\documentclass[aps,prl,twocolumn,showpacs,groupedaddress,floatfix]{revtex4}
\usepackage{graphicx}
\usepackage{dcolumn}
\usepackage{bm}
\usepackage{amssymb}
\usepackage{epsfig}
\usepackage{amsmath}
\usepackage{amsfonts}
\usepackage{feynmf}
\providecommand{\U}[1]{\protect\rule{.1in}{.1in}}

\newcommand{\met}{\mbox{\ensuremath{\slash\kern-.7emE_{T}}}}
\newcommand{\mht}{\mbox{\ensuremath{\slash\kern-.7emH_{T}}}}
\begin{document}
\title{Search for scalar top quarks in the acoplanar charm jets and
missing transverse energy final state in $p\bar{p}$ collisions at $\sqrt{s}=1.96$ TeV }
%\date{\today}
\date{March 14, 2008}
%The following information is for internal review, please remove them for submission
%\leftline{Version 0.9 as of \today}
%\leftline{Primary authors: Mansoora Shamim, Tim Bolton}
%\leftline{To be submitted to PLB}
%\rightline{Comment to {\tt d0-run2eb-019@fnal.gov}} \rightline{by March 4, 2008}

%the following line is for submission, including submission to the arXiv!!
\hspace{5.2in} \mbox{FERMILAB-PUB-08-063-E}
% LIST_OF_AUTHORS_R2.TEX               2/19/08              
%
\author{V.M.~Abazov$^{36}$}
\author{B.~Abbott$^{75}$}
\author{M.~Abolins$^{65}$}
\author{B.S.~Acharya$^{29}$}
\author{M.~Adams$^{51}$}
\author{T.~Adams$^{49}$}
\author{E.~Aguilo$^{6}$}
\author{S.H.~Ahn$^{31}$}
\author{M.~Ahsan$^{59}$}
\author{G.D.~Alexeev$^{36}$}
% ...

\affiliation{\vspace{0.1 in}(The D\O\ Collaboration)\vspace{0.1 in}}
% ...
\affiliation{$^{6}$University of Alberta, Edmonton, Alberta, Canada,
                Simon Fraser University, Burnaby, British Columbia, Canada,
                York University, Toronto, Ontario, Canada, and
                McGill University, Montreal, Quebec, Canada}
% ...
\affiliation{$^{29}$Tata Institute of Fundamental Research, Mumbai, India}
% ...
\affiliation{$^{31}$Korea Detector Laboratory, Korea University, Seoul, Korea}
% ...
\affiliation{$^{36}$Joint Institute for Nuclear Research, Dubna, Russia}
% ...
\affiliation{$^{49}$Florida State University, Tallahassee, Florida 32306, USA}
% ...
\affiliation{$^{51}$University of Illinois at Chicago,
                Chicago, Illinois 60607, USA}
% ...
\affiliation{$^{59}$Kansas State University, Manhattan, Kansas 66506, USA}
% ...
\affiliation{$^{65}$Michigan State University,
                East Lansing, Michigan 48824, USA}
% ...
\affiliation{$^{75}$University of Oklahoma, Norman, Oklahoma 73019, USA}
% ...

%input Dzero author list
\begin{abstract}
\noindent We present a search for the pair production of scalar top quarks, $\tilde{t}$,
using $995$ pb$^{-1}$ of data collected in $p\bar{p}$ collisions
with the D0 detector at the Fermilab Tevatron Collider at $\sqrt{s} = 1.96$ TeV. Both scalar top quarks are assumed to decay into a charm quark and a neutralino ($\tilde{\chi}^{0}_{1}$), where $\tilde{\chi}^{0}_{1}$ is the lightest supersymmetric particle. This leads to a final state with two acoplanar charm jets and missing transverse energy.  We find the yield of such events to be consistent with the standard model expectation, and exclude sets of $\tilde{t}$ and $\tilde{\chi}^{0}_{1} $ masses at the $95\%$ C.L. that substantially extend the domain excluded by previous searches.
\end{abstract}

\maketitle
\begin{figure}[ptb]
\includegraphics[scale=0.35]{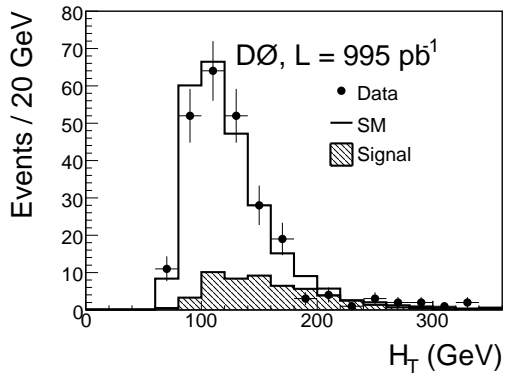}
\caption{Distributions of $H_{T}$
after applying optimized requirements on $\met$ and $\mathit{S} $ for data
(points with error bars), SM background (histogram), and a
signal with $m_{\tilde{t}}= 150$ GeV and $m_{\tilde{\chi}_{1}^{0}}= 70 $ GeV (hatched histogram). }
\label{ht}
\end{figure}
\begin{figure}[ptb]
\includegraphics[scale=0.35]{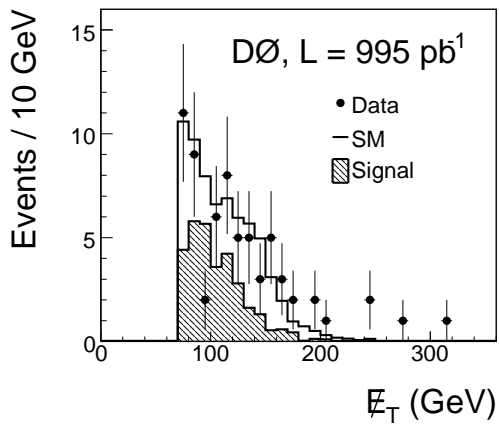}
\caption{Final distributions of $\met$ for data (points with error bars), SM background (histogram), and a signal with $m_{\tilde{t}}= 150$ GeV and $m_{\tilde{\chi}_{1}^{0}}= 70$ GeV (hatched histogram). }
\label{met}
\end{figure}
\begin{figure}[ptb]
\centering
\includegraphics[scale=0.4]{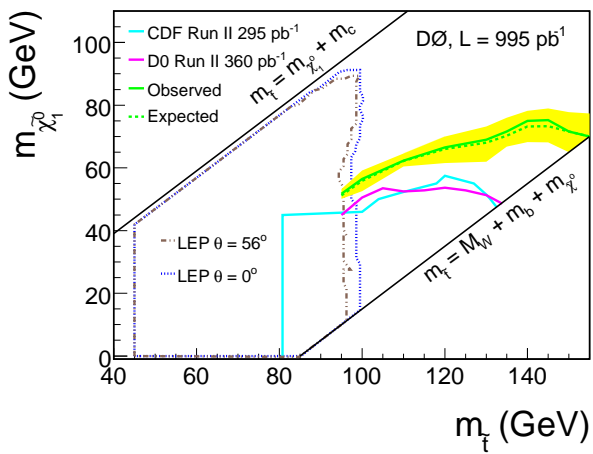}
\caption{Region in the $\tilde{t}$--$\tilde{\chi}_1^0$ mass
plane excluded at the $95\%$ C.L. by the present search. The observed
(expected) exclusion contour is shown as the green solid (dashed) line. The
yellow band represents the theoretical uncertainty on the scalar top quark pair production cross
section due to PDF and renormalization and factorization scale choice. Results from
previous searches~\cite{lep,cdf2,dzeroruntwo} are also shown. }
\label{contour}
\end{figure}

\indent Supersymmetry (SUSY) may provide a solution to the hierarchy problem if the SUSY particles have masses less than $1$ TeV, strongly motivating the searches for supersymmetric objects at the Fermilab Tevatron Collider. SUSY
predicts the existence of  partners with identical quantum numbers to all
standard model (SM) particles except for spin.
There exist two spin zero SUSY partners of the top quark corresponding to the
latter's left and right handed states. Several arguments exist in favor of a light scalar top quark ($\tilde{t}$).  The $\tilde{t}$ mass $m_{\tilde{t}}$ receives negative contributions proportional to the top quark Yukawa coupling in the renormalization group equations.  This makes the $\tilde{t}$ weak eigenstates lighter than other squarks~\cite{susy}.  Mixing between the left and right handed superpartners of the top quark, being proportional to the top quark mass $m_t$, leads to a large mass splitting between the two physical eigenstates.  This makes one of the $\tilde{t}$ considerably lighter than the other.    Additionally, a light $\tilde{t}$ that strongly couples to the Higgs boson could also generate a large enough CP violating phase to explain the mechanism for electroweak baryogenesis~\cite{baryog}. \newline\indent In R-parity conserving models~\cite{rparity}, the lightest supersymmetric particle (LSP) is stable, and cosmological constraints indicate that it should be neutral and colorless~\cite{cosmo}.  In the following we assume conservation of R-parity and take $\tilde{\chi}_{1}^{0} $, the lightest of four SUSY particles that result from the mixing between the SUSY partners of the SM neutral gauge and Higgs bosons, to be the LSP. 
\newline\indent In the search reported in this Letter, we consider the range $m_{\tilde{t}} < m_{b}+ m_{\tilde{\chi}_{1}^{+}}$ and  $m_{\tilde{t}}< m_{W}+m_{b}+m_{\tilde{\chi}_{1}^{0}}$, where $m_{b}$ is the $b$ quark mass, $m_{\tilde{\chi}_{1}^{0}}$ is the $\tilde{\chi}_{1}^{0}$ mass and $m_{\tilde{\chi}_{1}^{+}}$ is the $\tilde{\chi}_{1}^{+}$ mass, with $\tilde{\chi}_{1}^{+}$ being the lighter of two mass eigenstates resulting from the mixing of the SUSY partners of charged gauge and Higgs bosons.  The dominant $\tilde{t}$ decay mode in this model is the flavor changing process $\tilde{t}\rightarrow c\tilde{\chi}_{1}^{0}$ and is assumed to occur with $100\%$ branching fraction.  
The $\tilde{t}\rightarrow t \tilde{\chi}_1^0$ decay is kinematically forbidden over the $\tilde{t}$ mass range accessible in this search, and the tree level four-body decays $\tilde{t}\rightarrow bf\bar{f}' \tilde{\chi}_1^0$ can be neglected~\cite{fourbd}.  
\newline \indent In $p\bar{p}$ collisions, $\tilde{t}$ pairs are produced via quark-antiquark annihilation and gluon fusion.  The $\tilde{t}$ pair production cross section ($\sigma_{\tilde{t}\bar{\tilde{t}}}$) primarily depends on $m_{\tilde{t}}$, and a weak dependence on other SUSY parameters affects only the higher-order corrections.  At $\sqrt{s} = 1.96$ TeV which is the centre-of-mass energy available at the Fermilab Tevatron collider, $\sigma_{\tilde{t}\bar{\tilde{t}}}$ at next-to-leading-order (NLO), calculated with {\sc prospino}~\cite{pros1}, ranges from $15$ pb to $1$~pb for $100 \leq m_{\tilde{t}} \leq 160$ GeV.  These cross sections are calculated using {\sc cteq6.1m} parton distribution functions (PDFs)~\cite{cteq} and equal renormalization and factorization scales $\mu_{\mathrm{rf}}= m_{\tilde{t}}$.  A theoretical uncertainty of $\approx20\%$ is estimated due to scale and PDF choices.  The $\tilde{t}\bar{\tilde{t}}$ event topology consists of two
acoplanar charm jets and missing transverse energy ($\met$) from the neutralinos that escape detection.   Searches for $\tilde{t}$ pair production in the jets plus missing transverse energy mode have been reported by collaborations working at the CERN LEP collider~\cite{lep}, and the CDF~\cite{cdf,cdf2} and D0~\cite{dzerorunone,dzeroruntwo} collaborations. The highest excluded mass to date is $m_{\tilde{t}} < 134 $ GeV ($95\%$ C.L.) for $m_{\tilde{\chi}^0_1} = 48 $ GeV~\cite{dzeroruntwo}.  \newline\indent The $\tilde{t}$ search is performed in the data collected with the D0 detector during Run IIa of the Tevatron and corresponds to an integrated luminosity of $995 \pm 61$ pb$^{-1}$~\cite{d0lumi}.  
A detailed description of the D0 detector can be found in~\cite{d0det}. The
central tracking system consists of a silicon microstrip tracker and a fiber
tracker, both located within a $2$ T superconducting solenoidal magnet. A
liquid-argon and uranium calorimeter covers pseudorapidity $|\eta|\lesssim 4.2$, where $\eta=-\ln[\tan(\theta/2)]$, and $\theta$ is the polar
angle with respect to the proton beam direction. An outer muon system,
covering $|\eta|<2$, consists of layers of tracking detectors and
scintillation counters on both sides of $1.8$ T iron toroids.
\newline\indent The data sample collected from April $2003$ to February $2006$ with the jets+$\met$ triggers was analyzed for the $\tilde{t}$ search.  The trigger conditions require the $ \mht $ and its separation from all jets to be greater than $30$~ GeV and $25^{\circ}$, respectively, where $ \mht $ is the transverse energy computed only from jets.  Jets are reconstructed using an iterative midpoint cone algorithm with radius ${\cal R}_{\mathrm {cone}} = 0.5$~\cite{d0jets}.  The data set is reduced to a sample of $1.5$ million events by requiring at least two jets with $p_T > 15$ GeV and $\met > 40$ GeV.  
\newline\indent 
Signal samples are simulated using {\sc pythia} $6.323$~\cite{pythia} for $m_{\tilde{t}}$  ranging from $95$ GeV to $165$ GeV and $\tilde{\chi}_{1}^{0}$ masses from $45$ GeV to $90$ GeV. The largest expected backgrounds for this search are $W$ and $Z$ bosons produced in association with jets, denoted as V+jets.  The V+jets and $t\bar{t}$ processes are simulated using {\sc alpgen} $2.05$~\cite{ALPG} interfaced with {\sc pythia} for the generation of initial and final state radiation and hadronization.  The background samples for single top quark and diboson production are simulated using {\sc comphep}~\cite{comphep} and {\sc pythia}, respectively.  The PDF set {\sc cteq6l1} is used for both signal and background samples, and all generated events are subjected to full {\sc geant}-based~\cite{geant} simulation of the detector response. Simulated signal and background events are
overlaid with recorded unbiased beam crossings to incorporate the effect of multiple interactions that occur in a single beam crossing.  After reconstruction, simulated events are weighted properly to ensure that the instantaneous luminosity distribution is the same in data and the simulated Monte Carlo (MC) samples.   A parametrization of the trigger efficiency measured from the
data is applied to simulated MC events in order to fold in trigger effects.  The multijet background, not included in the MC samples, is directly estimated from data.\newline%
\indent A large data sample of
$Z/\gamma^{\ast}(\rightarrow ee)$ + jets events, corresponding to an integrated luminosity of $1067 \pm 65$~ pb$^{-1}$, from the same data period as the $\tilde{t}$ search, is used to improve
the prediction of V+jets backgrounds.
For this study, $Z$ boson candidates are selected using two high transverse
energy ($E_{T}>15$ GeV) clusters that deposit more than $90\%$
of their energy in the electromagnetic calorimeter, have shower shapes
consistent with expectations for electrons, are matched with tracks
reconstructed in the central tracker, and form an invariant mass between $65$ GeV and $115$ GeV.  At least two jets with $p_{T} > 15$ GeV and $|\eta_{\mathrm{det}}| < 2.5$ are required, where $|\eta_{\mathrm{det}}|$ is the jet pseudorapidity calculated using the assumption that the jet originates from the detector center.  The predicted number of $Z/\gamma^{\ast}(\rightarrow ee) +\geq 2$ jets events is calculated using {\sc alpgen} after correcting for differences in electron
and jet reconstruction efficiencies between data and MC and normalizing the
MC to the inclusive number of $Z$ boson events in data.  The {\sc alpgen} prediction is corrected in each jet multiplicity bin by a reweighting function that depends on the transverse momentum of the $Z$ boson to obtain better agreement between the model and data.  The reweighting function is derived by fitting the ratio of the transverse momentum distribution of $Z$ boson data to that from the MC prediction.  After reweighting, all other kinematical variables in the $Z/\gamma^{\ast}(\rightarrow ee) + \geq 2 $ jets sample applicable to the $\tilde{t}$ search are well described by MC.
\newline\indent The multijet background in $Z/\gamma^{\ast}(\rightarrow ee)$ + jets events is estimated from a fit to the dielectron invariant mass distribution.  The ratio of the number of $ee$ events produced by $\gamma^{\ast}$ to the number of $Z + \gamma^{\ast}$  events is
determined from MC and used to extract the multijet contribution by fitting the
dielectron invariant mass in data with an exponential function for the multijet+$\gamma^{\ast}$
contribution and a Breit-Wigner convolved with a Gaussian for $Z$ boson events.  
\newline\indent For the $\tilde{t}$ search, the predicted SM background from V+jets sources is normalized to the number of  $Z/\gamma^{\ast}(\rightarrow ee) + 2$ jets events after subtracting the multijet background, $ N_{Z(ee)+2}^{\mathrm{data}}$.  As an example, the normalization weight assigned to simulated  $Z(\rightarrow\nu\bar{\nu})$ events with $n$ light partons is
\begin{equation}
w_{\mathrm{MC}}^{Z(\nu\bar{\nu})+ n} = f\frac{N_{Z(ee)+2}^{\mathrm{data}}}{N_{Z(\nu\bar{\nu})+ n}^{\mathrm{MC}}}\frac{\sigma_{Z(\nu\bar{\nu})+n}^{\mathrm{ALP}}}{\sigma_{Z(ee)+2\mathrm{lp}}^{\mathrm{ALP}}}\frac{\epsilon_{Z(\nu\bar{\nu})+n}}{\epsilon_{Z(ee)+2\mathrm{lp}}}.\label{eq1}
\end{equation}
Here $N_{Z(\nu\bar{\nu})+n}^{\mathrm{MC}}$ is the number of simulated $Z(\rightarrow\nu\bar{\nu})+$ $n$ light parton jets events; $\sigma_{Z(\nu\bar{\nu})+ n}^{\mathrm{ALP}}$ and $\sigma_{Z(ee)+2\mathrm{lp}}^{\mathrm{ALP}}$ are the cross sections predicted by {\sc alpgen} for $Z(\rightarrow\nu\bar{\nu})+$ $n$ and $Z/\gamma^\ast(\rightarrow ee)+$ $2$ light parton jets, respectively; and $\epsilon_{Z(\nu\bar{\nu})+ n}$ and $\epsilon_{Z(ee)+2\mathrm{lp}}$ are the corresponding detection efficiencies.  The factor $f=0.89\pm 0.02$ is applied to correct for three effects: the absence of $\gamma^{\ast}$ contribution to $Z(\nu\bar{\nu})+$ jets events, the normalization of MC light jets to a data sample that contains all flavors of jets, and the difference in the luminosities of the data set used for the $\tilde{t}$ search ($995$ pb$^{-1}$) and the $Z/\gamma^\ast(\rightarrow ee) + 2$ jets data set ($1067$ pb$^{-1}$). 
\newline\indent
The normalization weight assigned to simulated  $W(\rightarrow \ell\nu)$+$n$ light partons is
\begin{equation}
\label{eq2}
w_{\mathrm{MC}}^{W(\ell\nu)+ n} = f\frac{N_{Z(ee)+2}^{\mathrm{data}}}{N_{W(\ell\nu)+ n}^{\mathrm{MC}}}\frac{\sigma_{W(\ell\nu)+n}^{\mathrm{ALP}}}{\sigma_{Z(ee)+2\mathrm{lp}}^{\mathrm{ALP}}}\frac{\epsilon_{W(\ell\nu)+n}}{\epsilon_{Z(ee)+2\mathrm{lp}}}\alpha(p_T), 
\end{equation}
where 
\begin{equation}
\alpha(p_T) = \frac{ \left[\frac{1}{\sigma_W^{\mathrm{NLO}}} \frac{d\sigma_W^{\mathrm{NLO}}}{dp_{T}}\right]  } { \left[ \frac{1}{\sigma_Z^{\mathrm{NLO}}} \frac{d\sigma_Z^{\mathrm{NLO}}}{dp_{T}} \right]}   \frac{ \left[\frac{1}{\sigma_Z^{\mathrm{ALP}}} \frac{d\sigma_Z^{\mathrm{ALP}}}{dp_{T}}\right]  } { \left[ \frac{1}{\sigma_W^{\mathrm{ALP}}} \frac{d\sigma_W^{\mathrm{ALP}}}{dp_{T}} \right]} ,
\end{equation}
is the product of the ratio of the normalized differential cross sections for $W$ and $Z$ bosons production at NLO~\cite{wpt} and predicted by {\sc alpgen}, respectively. 
\newline\indent The motivation behind using this technique is to
lower the luminosity times cross section uncertainty ($\approx 6.1 \% \oplus 15\%$) on the predicted number of events towards the $5\%$ statistical uncertainty of the $Z/\gamma^{\ast}(\rightarrow ee)+2$ jets normalization sample.  The combined $15\%$ uncertainty on the theoretical cross section  for various background processes is mainly due to the  choice of PDF and the renormalization and factorization scale.  The signal and smaller backgrounds such as $t\bar{t}$, diboson, and single top quark production are normalized using the measured absolute luminosity.  For these processes NLO cross sections were computed with {\sc mcfm} $5.1$~\cite{mcfm}.
\begin{table}
\caption{Numbers of data events and cumulative signal efficiency for $ m_{\tilde{t}}=150$ and $m_{\tilde{\chi}_{1}^{0}}= 70 $ GeV after each event selection.}
\begin{center}
\begin{ruledtabular}
\begin{tabular}{lrr}
Selection& Events left&Signal eff. ($\%$)\\
\hline
Initial selection and trigger   &$       1.5\times10^{6}$ &$ 55.9$\\
{\bf C1}: exactly two jets&$ 464477$ &$ 29.5$\\
{\bf C2}: $\mht > 40$ GeV&$ 440161$ &$ 27.5$\\
{\bf C3}: $\Delta \phi (\mathrm{jet}_1, \mathrm{jet}_2) < 165^\circ$ &$278505$&$26.5$\\
{\bf C4}: jet-1 $p_T> 40$ GeV &$216382$&$ 24.7$\\
{\bf C5}: jet-1 $|\eta_{\mathrm{det}}|< 1.5 $ &$ 113591$&$24.6$\\
{\bf C6}: jet-2 $p_T > 20$ GeV &$   80987$ &$22.0$\\
{\bf C7}: jet-2 $|\eta_{\mathrm{det}}|< 1.5$ &$ 62910$&$20.1$\\
{\bf C8}: jet-1 jet-2 CPF $ > 0.85$ &$49140$&$19.8$\\
{\bf C9}: isolated track veto  &$23832$&$13.4$\\
{\bf C10}: isolated electron veto &$ 23194$&$13.3$\\
{\bf C11}: isolated muon veto &$23081$&$13.3$\\
{\bf C12}: $\Delta \phi_{\mathrm{max}} - \Delta\phi_{\mathrm{min}} < 120^\circ$ &$9753$&$12.6$\\
{\bf C13}: {\it A} $ > -~ 0.05 $&$3733$&$12.0$\\
{\bf C14}: $\Delta \phi (\mathrm{jet},~ \met ) >  50^\circ$ &$3375$&$11.6$\\
{\bf C15}: $\met > 60 $ GeV &$2288$&$10.0$\\
\end{tabular}
\end{ruledtabular}
\end{center}
\label{tab:table1}
\end{table}
\newline\begin{figure*}[ptb]	
\includegraphics[scale=0.55]{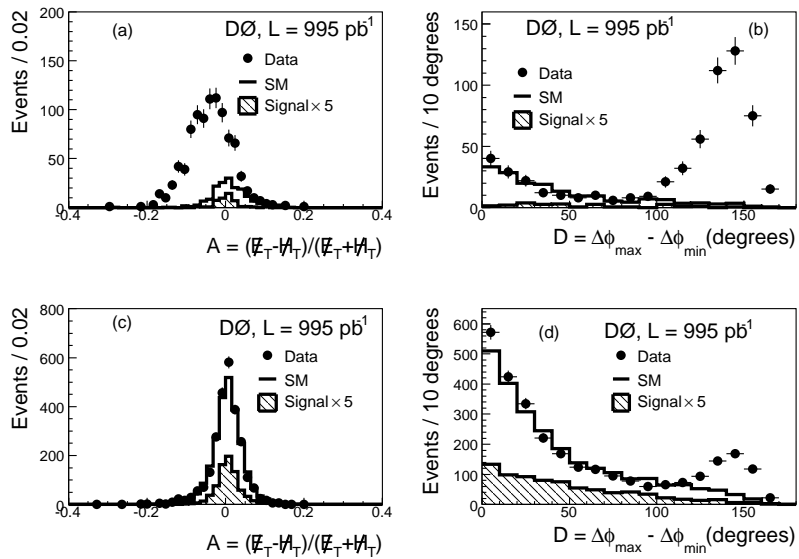}	
\caption{Distributions of the asymmetry ${\it A} = (~\met-\mht) / (~\met+\mht) $ with the requirement on $\mathit{D} = \Delta \phi _{\mathrm{max}} - \Delta\phi _{\mathrm{min}}$ inverted (a) and applied (c).  Distributions of $\mathit{D}$ with the requirement on ${\it A}$ inverted (b) and applied (d) for data (points with error bars), for SM backgrounds (histogram), and for a signal with $m_{\tilde{t}} = 150$ GeV and $m_{\tilde{\chi}_1^0} = 70$ GeV (hatched histogram).  In all plots the signal contribution has been scaled up by five and $\met > 60$ GeV is required.  The excess in data at ${\it A}=0$ and $\mathit{D} = 0-10$ degrees is consistent with the systematic uncertainties on the predicted background.  }
\label{epsartt}	
\end{figure*}
The search strategy for $\tilde{t}$ involves three steps which include the application of the selection criteria on kinematical variables, heavy flavor (HF) tagging and optimization of the final selection depending on $\tilde{t}$ and $\tilde{\chi}_1^0$ masses.  The data set for the $\tilde{t}$ search is reduced to a sample of $2288$ potential $\tilde{t}\bar{\tilde{t}}$ candidates, by applying the $15$ selection criteria denoted by $\mathbf{C1-C15}$ and summarized in Table \ref{tab:table1}.
The main motivation for $\mathbf{C1}$ is to reduce the multijet background.  Requirements $\mathbf{C2}$ to $\mathbf{C7}$ help in reducing the $W+$jets and
multijet backgrounds.  The condition on the charged particle fraction (CPF) in $\mathbf{C8}$ requires that at least $85\%$ of the jets' charged particle transverse momenta be
associated with tracks originating from the selected primary vertex in the event. This track confirmation requirement removes events with spurious $ \met $ due to the choice of an incorrect primary vertex.  $\mathbf{C9-C11}$ are applied to reject $W+$jets
background with isolated leptons from $W$ boson decay. For an electron to be
isolated, the energy deposited in the calorimeter in a cone of radius $0.4$ in $\eta-\phi$ around the electron direction should not be more than $15\%$ of the energy deposited in the electromagnetic layers inside a cone of radius $0.2$. A muon is declared isolated if the sum of the energies of all tracks other than the muon in a cone of radius $0.5$ is less
than $2.5$ GeV, and the calorimeter energy deposited in a hollow cone with inner and outer
radii $0.1$ and $0.4$ around the muon direction is less than $5$ GeV. A track with $p_{T}> 5$
GeV is considered isolated if no other track with $p_{T}> 1.5$ GeV is found
in a hollow cone of inner and outer radii $0.1$ and $0.4$ around the track considered. This condition also helps suppress
backgrounds with $\tau$ leptons where the $\tau$ decays hadronically.  Remaining instrumental background is removed using a quantity defined by the
angular separation between all jets and the $\met$ of the event,
$\mathit{D}=\Delta\phi_{\max} - \Delta\phi_{\min}$, where $\Delta\phi_{\max
}\left(  \Delta\phi_{\min}\right)  $ is the largest (smallest) azimuthal
separation between a jet and $\met$; and an asymmetry variable
defined as $\mathit{A}=\left(\met-\mht\right)
/\left(\met+\mht\right)  $.  The requirements applied on these variables are given by $\mathbf{C12}$ and $\mathbf{C13}$.  Figure \ref{epsartt} shows
that both of these variables are very effective in eliminating multijet
background which dominates in data for large $\mathit{D}$ and negative $\mathit{A}$. \\
\indent The $2288$ events selected in data can be compared to the $2199 \pm 18^{+316}_{-321}$ events predicted from the simulation normalized to $Z/\gamma^\ast(\rightarrow ee) + 2$ jets events or $2292 \pm 19^{+527}_{-532}$ events predicted using absolute luminosity
normalization, with the first quoted uncertainty due to finite MC
statistics and the second due to systematic effects described in more detail below.  The small remaining multijet background in the $\tilde{t}$ search analysis is estimated after applying all analysis conditions shown in Table \ref{tab:table1} except that on $\met$.   The background subtracted $\met$ distribution is fitted in the control region ($ 40 \leq \met \leq 60 $ GeV) with exponential and power law functions, and the estimated contribution is extrapolated into the signal region ($\met > 60$ GeV).  The average of the two results is taken as the multijet background estimate, while the difference between the two fit results is taken as the systematic uncertainty.  This amounts to $ 14.4 \pm 10.7 $ (stat) $ \pm $ $ 5.1 $ (sys) events contributed by  multijet background before HF tagging and optimization of selection cuts.
\newline\indent
After selecting candidate events on the basis
of topology, HF tagging is used to identify charm jets in the final
state. A neural network (NN) tagging tool~\cite{nntag} that combines information from
three different D0 HF taggers to maximize the $b$ quark tagging efficiency ($\approx 73\%$) is used for this purpose.  \\
\indent The first tagger converts information from the impact parameter of the tracks identified in a jet into a
probability that all tracks originate from the primary vertex, where the impact
parameter is the distance of closest approach to the interaction
point in a plane perpendicular to the beam axis.  The second tagger identifies the presence of vertices that are significantly displaced from the primary vertex and associated with a jet. The
third tagger makes use of the number of tracks with large impact parameter
significance, where the significance is defined as the ratio of the impact parameter to its uncertainty.  \ The result of the combination is a NN output.  A requirement on the NN output is made that preserves high efficiency for detection of charm jets ($\approx 30\%$) with a  $\approx6\%$ probability for a light parton jet to be mistakenly tagged.   The efficiency for $c$ jet tagging is obtained by scaling the $b$ jet tagging efficiency measured in the data by the $c$-tagging-to-$b$-tagging efficiency ratio computed in the MC.
\begin{figure}[ptb]
\includegraphics[scale=0.35]{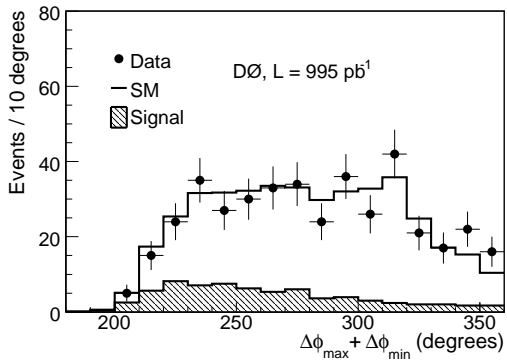}\caption{Distributions of
$\mathit{S} = \Delta\phi_{\mathrm{max}} + \Delta\phi_{\mathrm{min}}$ for data (points with error bars), SM background (histogram), and a signal with $m_{\tilde{t}}= 150$ GeV and $m_{\tilde{\chi}_{1}^{0}}= 70$ GeV (hatched histogram) after requiring HF tagging but before optimization. }
\label{dphi}
\end{figure}
\newline \indent At the final stage of the analysis, additional selection criteria on three kinematic variables; $\met$, $\mathit{S}=\Delta\phi_{\mathrm{max}} + \Delta\phi_{\mathrm{min}}$, and $H_{T}$, with $H_{T}$ defined as the scalar sum of the $p_{T}$ of all jets, are optimized by maximizing the expected lower limit on the neutralino mass for a given $m_{\tilde{t}}$.  The variable $\mathit{S}$ after requiring at least one jet in the event to be HF tagged is shown in Fig. \ref{dphi}. \newline \indent Minimum values of $H_{T}$ are varied from $60$ GeV to $140$~ GeV in steps of $20$ GeV, while those for $\met$ are varied from $60$ GeV to $100$ GeV in steps of $10$ GeV.  Events having the values of these quantities above the minima are kept.
Maximum values of $\mathit{S}$ are tested between $240^{\circ}$ and $320^{\circ}$ in steps of $20^{\circ}$, and events having $\mathit{S}$ below the minimum are retained.    
\begin{table}[ptb]
\caption{Optimized values of selections, numbers of observed data and predicted background events.  A requirement of $\met > 70$ GeV was chosen in all cases.
The values of $m_{\tilde{t}}$ and $H_{T}$ are in GeV while those for $\mathit{S}$ are in degrees.
}
\renewcommand{\arraystretch}{2.0}
\begin{center}
\begin{ruledtabular}
\begin{tabular}
{cclclcc}
$m_{\tilde{t}}$ ~~~~& &~~~~$H_{T} $& & ~~~~$\mathit{S} $ & Observed & Predicted\\
\hline
$95-130$ & &$> 100$ & &$< 260$ & $83 $ & $85.3\pm 1.8^{+ 12.8}_{-13.0}$\\
$135-145 $ & & $> 140$ & &$< 300$ & $57 $ & $59.0\pm1.6^{+ 8.5}_{-8.8}$\\
$150-160$ & & $> 140$ & &$< 320$ & $66 $ & $66.6\pm1.1^{+ 9.6}_{-10.0}$\\
\end{tabular}
\end{ruledtabular}
\end{center}
\label{tabmult}
\end{table}
\begin{table}
\caption{For three $\tilde{t}$ and $\tilde{\chi}_1^0$ mass combinations: signal efficiencies and the numbers of signal events expected.  The first errors are statistical and second systematic.  The nominal (NLO) signal cross section and upper limit at the $95\%$ C.L. are also shown. }
\renewcommand{\arraystretch}{2.0}
\begin{center}
\begin{ruledtabular}
\begin{tabular}{ccccccc}
($m_{\tilde{t}}$, $m_{\tilde{\chi}^0_1}$) & Efficiency & Expected Signal&& $\sigma_{\mathrm{nom}}$ && $\sigma_{\mathrm{95}}$\\
GeV&($\%$)&Events &&pb&& pb\\
\hline
$(130, 55)$&$1.5$	&$51.9$	$\pm$	$2.7^{+7.2}_{-7.1}$	&&$3.44$&&$2.41$\\
$(140, 80)$&$0.9$	&$19.6$	$\pm$	$0.8^{+2.8}_{-2.5}$	&&$2.24$&&$2.87$\\
$(150, 70)$&$2.1$	&$30.8$	$\pm$	$1.2^{+4.2}_{-3.7}$     &&$1.49$&&$1.42$	\\
\end{tabular}
\end{ruledtabular}
\label{sigmaul}
\end{center}
\end{table}
For each set of requirements, the expected value of the signal confidence
level $\langle CL_{s}\rangle$~\cite{cls} under the hypothesis that only
background is present is evaluated using all $\tilde{t}$ and $\tilde{\chi}_1^0$ mass
combinations, taking into account systematic uncertainties.   The set of
criteria that return $\langle CL_{s}\rangle = 0.05$ for the highest
neutralino mass corresponding to a given $m_{\tilde{t}}$ are chosen to be the optimal ones.  \newline \indent The optimized values of the selections for different $ m_{\tilde{t}}$ are given
in Table \ref{tabmult} along with the number of events observed in data and
expected SM background. In all cases a requirement of $\met\geq 70$ GeV is
imposed.  No contamination remains from multijet background at this point in the analysis; it is therefore neglected while setting the limit.  Efficiencies for three signal mass points along with the expected numbers of events are shown in Table \ref{sigmaul}.  The distribution of $H_T$ after optimization but with the constraint on $H_T$ removed is shown in Fig. \ref{ht}. The final distribution of   $\met$ is shown in Fig. \ref{met}.  The detailed SM background composition is given in Table \ref{stop160}.\\
\begin{table}[ptb]
\caption{Numbers of predicted background events from different SM sources for a selection optimized for $m_{\tilde{t}}\geq 150$~ GeV.  The uncertainties are due to the limited MC statistics.}
\begin{center}
\begin{ruledtabular}
\begin{tabular}{cccc}
\multicolumn{1}{l}{SM process} &\multicolumn{3}{r}{Number of events}\\
\hline
\multicolumn{1}{l}{$W(\rightarrow \ell \nu)$ + jets} && & $20.0 $ $\pm$ $ 0.7 $\\
\multicolumn{1}{l}{$Z(\rightarrow\nu\bar{\nu})$ + jets} && & $15.8 $ $\pm$  $0.5 $\\
\multicolumn{1}{l}{$W(\rightarrow \ell\nu)$ + HF $(b\bar{b}, c\bar{c}) $ }&& & $12.6 $
$\pm$ $0.5$\\
\multicolumn{1}{l}{$Z(\rightarrow\nu\bar{\nu})$ + HF $(b\bar{b}, c\bar{c})$ } &&&  $11.6 $ $\pm$  $0.4 $\\
\multicolumn{1}{l}{$t\bar{t} $ and single top  }& && $~3.7 $ $\pm$  $0.1 $\\
\multicolumn{1}{l}{$WW, WZ, ZZ$  }&& & $~2.7 $ $\pm$  $0.1 $\\
\multicolumn{1}{l}{$Z(\rightarrow \ell\ell)$ ($e,\mu,\tau) $ + jets } && & ~~$0.1 $ $\pm$  $0.01 $\\
\multicolumn{1}{l}{$Z(\rightarrow \ell\ell)$ ($e,\mu,\tau)$ + HF $( b\bar{b}, c\bar{c} )$  }&& & ~~$0.1 $ $\pm$  $0.01 $\\
\hline
\multicolumn{1}{l}{Total} &\multicolumn{1}{l}{}&&$66.6\pm 1.1$\\
\end{tabular}
\end{ruledtabular}
\end{center}
\label{stop160}%
\end{table}
Systematic uncertainties are evaluated for each $\tilde{t}$ and $\tilde{\chi}_1^0$ mass combination for the optimized set of requirements. Sources of
systematic uncertainty include jet energy scale, jet energy resolution,
jet identification and reconstruction, the jet multiplicity requirement, trigger efficiency, data to MC scale
factors, normalization of background, HF tagging, luminosity determination, choice of PDF, and $W$ boson $p_T$ reweighting.  The
effect of the jet multiplicity requirement on the background is studied using $Z/\gamma^\ast(\rightarrow ee)$ +
jets events. The spectrum of transverse momentum of the
third jet in data events with three or more jets is observed to be very well
described by the simulation generated with {\sc alpgen}. The $\approx1\%$ statistical uncertainty of the lowest $p_T$ bin, where the bulk of the events are, is taken as a systematic uncertainty introduced by the jet
multiplicity requirement. \ To study the effect of the same requirement on the $\tilde{t}$ signal,
where a third jet enters an event primarily through initial or final state radiation, the $p_{T}$ spectrum of the leading jet in simulated $Z/\gamma^{\ast}(\rightarrow ee)$ events generated with {\sc pythia} is examined. Comparison between data and
simulation shows a slight excess in data in the low $p_T$ bin; this
discrepancy is used to estimate a systematic uncertainty of $\pm 1.5\%$ on the
signal acceptance attributable to the jet multiplicity requirement.   The uncertainty on the signal acceptance and background estimation due to the PDF choice was determined using the {\sc cteq6.1m} PDF set.  \newline \indent The combined $10\%$ uncertainty on the background normalization includes: $5\%$ uncertainty from $Z/\gamma^{\ast}(\rightarrow ee)$ + jets statistics assigned to all V+jets samples; $50\%$ uncertainty on the NLO cross section assigned to the V + HF background;  $6.1\%$ luminosity uncertainty assigned to $t\bar{t}$, diboson, and single top quark background; and $8\%, 6\% $ and $15\%$ uncertainties on NLO cross sections for $t\bar{t}$, diboson, and single top quark production, respectively.  The uncertainty on the background estimation due to the $W$  boson $p_T$ reweighting is estimated using two different methods to estimate the $W$+jets background.  In the first method, the $W$+jets background is estimated using the expression given in Eq. \ref{eq2}.  In the second method, the same reweighting function as applied to the $Z$ boson was used to reweight the $W$ boson $p_T$ which is equivalent to setting $\alpha(p_T) = 1$ in Eq. \ref{eq2}.  Detailed estimates of all systematic uncertainties are given in Table \ref{uncerttag}.
\begin{table}[ptb]
\caption{Breakdown of systematic uncertainties on the SM background and for
a signal point with $m_{\tilde{t}} = 150$ GeV and $m_{\tilde{\chi}_{1}^{0}} = 70$ GeV.}
\begin{ruledtabular}
\begin{center}
\begin{tabular}{lllll}
Source & & SM  && Signal\\
 &&background&&\\
\hline
Jet energy & & $+1.7\%$ && $+2\%$\\
 scale&&$-2.5\%$ && $-4\%$\\
Jet resolution & & $\pm 1\% $&& $\pm 1\%$ \\
Jet reconstruction &  &$\pm0.8\% $&& $\pm 0.1 \% $\\
and identification &  &&& \\
Trigger & & $\pm  6 \%$&& $\pm 6 \%$\\
Scale factor & & $\pm 5 \%$&& $\pm 5 \%$\\
Normalization  & & $\pm 10\% $& &--\\
Luminosity  && -- && $\pm6.1\%$\\
HF  tagging&&  $\pm 4.1\%$ && $\pm 3.5\%$\\
PDF choice  && $ \pm 4 \%$ &&$+8.7\%$\\
 && -- &&$-5.5\%$\\
Two jet cut  && $\pm0.9\%$ && $\pm1.5\%$\\
$W$ boson $p_T$ && $\pm 3\%$ &&--\\
reweighting && && 
\end{tabular}
\end{center}
\end{ruledtabular}
\label{uncerttag}
\end{table}
\newline\indent Using the assumption that $\tilde{t}$ decays into a charm quark and a neutralino with 100$\%$ branching fraction and the nominal $\tilde{t}$ pair production cross
section, the largest $m_{\tilde{t}}$ excluded by this analysis is $155$ GeV, for a
neutralino mass of $70$ GeV at the $95\%$ C.L.  With the theoretical uncertainty on the $\tilde{t}$ pair production cross section taken into account, the largest limit on $m_{\tilde{t}}$ is $150$ GeV, for $m_{\tilde{\chi}_{1}^{0}} = 65$~ GeV.  These results are shown in Fig. \ref{contour}.  \newline \indent In summary, D0 has searched for scalar top quarks in jets plus missing transverse energy final states using $1$ fb$^{-1}$ of data.  No evidence for $\tilde{t}$ production has been found.  This analysis substantially extends the excluded region of the $\tilde{t}$ -- $\tilde{\chi}_1^0 $ mass plane over the searches carried out previously. 
\newline 
% acknowledgement_paragraph_r2.tex                         2/19/08
%
We thank the staffs at Fermilab and collaborating institutions, 
and acknowledge support from the 
DOE and NSF (USA);
CEA and CNRS/IN2P3 (France);
FASI, Rosatom and RFBR (Russia);
CNPq, FAPERJ, FAPESP and FUNDUNESP (Brazil);
DAE and DST (India);
Colciencias (Colombia);
CONACyT (Mexico);
KRF and KOSEF (Korea);
CONICET and UBACyT (Argentina);
FOM (The Netherlands);
STFC (United Kingdom);
MSMT and GACR (Czech Republic);
CRC Program, CFI, NSERC and WestGrid Project (Canada);
BMBF and DFG (Germany);
SFI (Ireland);
The Swedish Research Council (Sweden);
CAS and CNSF (China);
and the
Alexander von Humboldt Foundation.

\clearpage
\end{document}